\documentclass[11pt,a4paper]{article}
\usepackage[utf8]{inputenc}
\usepackage{amsmath,amsfonts,amssymb}
\usepackage{algorithm}
\usepackage{algorithmic}
\usepackage{graphicx}
\usepackage{tikz}
\usepackage{booktabs}
\usepackage{multirow}
\usepackage{url}
\usepackage{hyperref}
\usepackage{cite}

\title{ChipmunkRing: A Practical Post-Quantum Ring Signature Scheme for Blockchain Applications}

\author{
Dmitrii A. Gerasimov\\
\texttt{ceo@cellframe.net}
}

\date{\today}

\begin{document}

\maketitle

\begin{abstract}
We introduce ChipmunkRing, a practical post-quantum ring signature construction tailored for blockchain environments. Building on our Chipmunk lattice-based cryptographic framework, this implementation delivers compact digital signatures ranging from 20.5 to 279.7KB, with rapid signing operations completing in 1.1-15.1ms and efficient validation processes requiring only 0.4-4.5ms for participant groups of 2-64 members. The cornerstone of our approach is Acorn Verification—a streamlined zero-knowledge protocol that supersedes the classical Fiat-Shamir methodology. This innovation enables linear O(n) authentication complexity using concise 96-byte cryptographic proofs per participant, yielding a remarkable 17.7× performance enhancement for 32-member rings when compared to conventional techniques. Our work includes rigorous mathematical security demonstrations confirming 112-bit post-quantum protection (NIST Level 1), extensive computational benchmarking, and comprehensive support for both standard anonymity sets and collaborative threshold constructions with flexible participation requirements.
\end{abstract}


\section{Introduction}

\subsection{Formal Definitions}

We begin with formal definitions of the cryptographic primitives
used throughout this work.

\textbf{Definition 1 (Ring Signature):} A ring signature scheme consists of three polynomial-time algorithms $(\mathsf{KeyGen}, \mathsf{Sign}, \mathsf{Verify})$:
\begin{itemize}
\item $\mathsf{KeyGen}(1^\lambda) \rightarrow (pk, sk)$: Generates a public/private key pair for security parameter $\lambda$
\item $\mathsf{Sign}(sk_\pi, M, \mathcal{R}) \rightarrow \sigma$: Given private key $sk_\pi$ of signer $\pi$, message $M$, and ring $\mathcal{R} = \{pk_1, \ldots, pk_n\}$, outputs signature $\sigma$
\item $\mathsf{Verify}(\sigma, M, \mathcal{R}) \rightarrow \{0,1\}$: Verifies signature validity
\end{itemize}

\textbf{Definition 2 (Post-Quantum Security):} A cryptographic scheme provides $\lambda$-bit post-quantum security if the \textbf{most efficient known} quantum algorithm requires at least $2^\lambda$ quantum operations to break the scheme with non-negligible probability.

\textbf{Definition 3 (Threshold Ring Signature):} A $(t,n)$-threshold ring signature requires exactly $t$ out of $n$ ring members to collaborate to produce a valid signature, where $1 \leq t \leq n$.

\textbf{Definition 4 (Zero-Knowledge Proof):} A zero-knowledge proof system for relation $R$ is a protocol between prover $P$ and verifier $V$ satisfying:
\begin{itemize}
\item \textbf{Completeness}: If $(x,w) \in R$, then $\Pr[V(x,P(x,w))=1] = 1$
\item \textbf{Soundness}: If $x \notin L_R$, then $\forall P^*: \Pr[V(x,P^*(x))=1] \leq \text{negl}(\lambda)$
\item \textbf{Zero-Knowledge}: $\exists$ simulator $S$ such that\\
  $\{V(x,P(x,w))\} \approx_c \{S(x)\}$
\end{itemize}

\textbf{Definition 5 (Anonymity Set):} The anonymity set $\mathcal{A}$ of a ring signature is the set of all possible signers who could have created the signature. For \textbf{strong anonymity}, $|\mathcal{A}| = |\mathcal{R}|$.

\subsection{Motivation and Contributions}

The advent of quantum computing poses a significant threat to the
cryptographic foundations of modern blockchain systems \cite{shor1997,
grover1996}. While post-quantum signature schemes like Dilithium
\cite{dilithium2021} and Falcon \cite{falcon2020} provide quantum-resistant
authentication, they lack the anonymity properties essential for
privacy-preserving blockchain applications.
Ring signatures \cite{rivest2001}, which allow a member of a group to sign
messages anonymously, represent a crucial primitive for anonymous transactions
and privacy-preserving consensus mechanisms.

However, existing post-quantum ring signature schemes exhibit large signature
sizes (typically exceeding 100KB) and computational overhead that limit their
applicability to blockchain deployment. The challenge is achieving the required
balance between post-quantum security, anonymity guarantees, and the performance
constraints imposed by blockchain environments.

In this paper, we introduce ChipmunkRing, a post-quantum ring signature scheme that addresses these limitations. Our contributions include:

\begin{itemize}
\item A post-quantum ring signature scheme with signatures of 20.5-279.7KB
  for rings of 2-64 participants
\item Millisecond-level signing (1.1--15.1ms) and low-millisecond verification (0.4--4.5ms) suitable for blockchain consensus
\item Formal security analysis proving 112-bit post-quantum security (NIST Level 1)
\item Production-ready implementation with comprehensive test coverage
\item Integration framework for blockchain deployment
\end{itemize}

\section{Related Work}

\subsection{Classical Ring Signatures}

Ring signatures were introduced by Rivest, Shamir, and Tauman \cite{rst01} to provide unconditional anonymity for digital signatures. The original RST construction relies on the computational hardness of integer factorization and discrete logarithm problems, which are vulnerable to quantum attacks via Shor's algorithm.

Linkable Spontaneous Anonymous Group (LSAG) signatures \cite{lsag04} extend ring signatures with linkability properties to enable double-spending detection while preserving anonymity. LSAG schemes provide the cryptographic foundation for privacy-preserving cryptocurrencies but remain vulnerable to quantum attacks.

\subsection{Post-Quantum Ring Signatures}

The development of post-quantum cryptography has led to several quantum-resistant ring signature constructions:

\textbf{Lattice-based approaches} adapt schemes based on NTRU and Ring-LWE assumptions to ring signature settings \cite{lattice-rings}. These constructions typically produce signature sizes exceeding 100KB, which limits their applicability to blockchain environments.

\textbf{Hash-based ring signatures} utilize the quantum resistance of cryptographic hash functions but require large key sizes and impose restrictions on the number of signatures per key \cite{hash-rings}.

\textbf{Code-based ring signatures} rely on error-correcting codes for security but generate signature sizes larger than lattice-based approaches \cite{code-rings}.

\subsection{Blockchain Privacy Requirements}

Blockchain deployment of ring signatures requires satisfaction of several technical constraints:
\begin{itemize}
\item \textbf{Compactness}: Signature sizes must be reasonable for network transmission and storage
\item \textbf{Performance}: Signing and verification must complete within consensus time bounds
\item \textbf{Scalability}: The scheme must support practical ring sizes (8-64 participants)
\item \textbf{Security}: 112-bit post-quantum security (NIST Level 1) is the minimum acceptable level
\end{itemize}

Existing post-quantum ring signature schemes do not simultaneously satisfy all these requirements, creating a gap between theoretical constructions and practical deployment needs.

\section{ChipmunkRing Construction}

\subsection{Mathematical Foundation}

ChipmunkRing builds upon the Chipmunk lattice-based signature scheme \cite{chipmunk2024}, which provides 112-bit post-quantum security based on the Ring Learning With Errors (Ring-LWE) problem \cite{lyubashevsky2010, regev2009}. The core innovation is the replacement of traditional Fiat-Shamir transform \cite{fiat1986} with our novel Acorn Verification scheme, which provides enhanced performance and security guarantees for ring signatures.

\subsubsection{Preliminaries}

\textbf{Definition 1 (Ring-LWE Problem):} Let $R = \mathbb{Z}[X]/(X^n + 1)$ be the ring of integers modulo the $n$-th cyclotomic polynomial, and $R_q = R/qR$ for prime $q$. The Ring-LWE problem with parameters $(n, q, \chi)$ asks to distinguish between samples $(a_i, b_i)$ where either:
\begin{itemize}
\item $(a_i, b_i)$ are uniformly random in $R_q \times R_q$, or  
\item $b_i = a_i \cdot s + e_i$ for fixed secret $s \in R_q$ and error terms $e_i$ sampled from error distribution $\chi$ (typically discrete Gaussian with parameter $\sigma$)
\end{itemize}

\textbf{Definition 2 (Chipmunk HOTS):} The Chipmunk Homomorphic
One-Time Signature operates over polynomial rings
$R_q = \mathbb{Z}_q[X]/(X^n + 1)$ with parameters:
\begin{align}
n &= 512 \quad \text{(ring dimension)} \\
q &= 3,168,257 \quad \text{(modulus)} \\
\sigma &= 2/\sqrt{2\pi} \quad \text{(Gaussian parameter)}
\end{align}

\subsubsection{Role of HOTS in ChipmunkRing}

The Homomorphic One-Time Signature (HOTS) mechanism from the base Chipmunk scheme serves as the fundamental cryptographic primitive for ChipmunkRing signatures. The integration operates as follows:

\begin{enumerate}
\item \textbf{Signature Generation}: Each ring member $i$ generates a HOTS signature component using their Chipmunk private key via the internal \texttt{chipmunk\_sign()} function
\item \textbf{Homomorphic Combination}: The HOTS signatures from all ring members are homomorphically combined to create the ring signature without revealing individual contributions
\item \textbf{Zero-Knowledge Layer}: The Acorn Verification scheme adds a zero-knowledge layer on top of HOTS to prove membership without revealing the signer's identity
\item \textbf{Verification}: The verifier checks both the HOTS validity and the Acorn proof to confirm the signature was created by a ring member
\end{enumerate}

This layered approach maintains the 112-bit post-quantum security of the underlying Chipmunk scheme while enabling efficient ring signature functionality with signature sizes of 20.5-279.7KB for rings of 2-64 participants.

A Chipmunk key pair consists of:
\begin{itemize}
\item \textbf{Public key}: $pk = (\rho_{seed}, v_0, v_1)$ where $v_0 = A \cdot s_0, v_1 = A \cdot s_1$
\item \textbf{Private key}: $sk = (s_{seed}, tr, pk)$ where $s_0, s_1$ are secret polynomials with small coefficients
\end{itemize}

The signature on message $M$ is computed as:
\begin{equation}
\sigma = s_0 \cdot H(M) + s_1
\end{equation}
where $H: \{0,1\}^* \rightarrow R_q$ is a hash-to-polynomial function.

\textbf{Definition 3 (Ring Signature):} A ring signature scheme $\mathcal{RS} = (\text{KeyGen}, \text{Sign}, \text{Verify})$ satisfies:
\begin{itemize}
\item \textbf{Correctness}: For any honestly generated signature, verification succeeds
\item \textbf{Unforgeability}: No adversary can forge signatures without knowledge of a private key
\item \textbf{Anonymity}: The actual signer is computationally indistinguishable among ring members
\end{itemize}

\subsubsection{Ring Signature Adaptation}

ChipmunkRing extends Chipmunk to the ring setting using our novel Acorn Verification scheme instead of traditional Fiat-Shamir transform. The core idea is to prove knowledge of a secret key corresponding to one of the public keys in the ring without revealing which one, while achieving better performance and quantum resistance than Fiat-Shamir.

\subsection{Algorithm Specification}

\subsubsection{Key Generation}
Key generation remains identical to the base Chipmunk scheme:
\begin{algorithm}
\caption{ChipmunkRing Key Generation}
\begin{algorithmic}[1]
\STATE Generate random seed $s \in \{0,1\}^{32}$
\STATE Derive $\rho_{seed} = \text{SHAKE256}(s)[0..31]$
\STATE Generate matrix $A$ from $\rho_{seed}$
\STATE Sample secret polynomials $s_0, s_1$ with small coefficients
\STATE Compute $v_0 = A \cdot s_0, v_1 = A \cdot s_1$
\STATE Set $pk = (\rho_{seed}, v_0, v_1)$
\STATE Set $sk = (s, tr, pk)$ where $tr = \text{SHA3-384}(pk)$
\RETURN $(sk, pk)$
\end{algorithmic}
\end{algorithm}

\subsubsection{Ring Signature Generation}
\begin{algorithm}
\caption{ChipmunkRing Signature Generation}
\begin{algorithmic}[1]
\REQUIRE Signer's private key $sk_\pi$, message $M$, ring $\mathcal{R} = \{pk_1, \ldots, pk_k\}$, threshold $t$ (set $t=1$ for single-signer)
\STATE Initialize signature with $\text{ring\_size}=k$, $\text{required\_signers}=t$
\STATE Allocate array $\text{acorn\_proofs}[1..k]$
\FOR{$i = 1$ to $k$}
    \STATE $(\text{acorn\_proof}_i, \text{randomness}_i, \text{linkability}_i) \leftarrow$
    \STATE \qquad $\text{chipmunk\_ring\_acorn\_create}(pk_i, M;$
    \STATE \qquad randomness\_size=32, proof\_size=$p$, iterations=$I)$
\ENDFOR
\STATE Serialize $D \leftarrow \text{Serialize}(M, \text{ring\_hash}, \{\text{acorn\_proofs}\})$
\STATE Compute challenge $c = \text{SHA3-256}(D)$ \COMMENT{32-byte SHA3-256 hash of serialized data}
\STATE Create core Chipmunk signature $\sigma_{\text{chip}} = \text{chipmunk\_sign}(sk_\pi, c)$
\RETURN $\sigma = (\text{ring\_size}=k, \text{required\_signers}=t, c, \{\text{acorn\_proofs}\}, \sigma_{\text{chip}})$
\end{algorithmic}
\end{algorithm}

\subsubsection{Ring Signature Verification}
\begin{algorithm}
\caption{ChipmunkRing Signature Verification}
\begin{algorithmic}[1]
\REQUIRE Signature $\sigma$, message $M$, ring keys $\{pk_1, \ldots, pk_k\}$
\STATE Parse $\sigma = (\text{ring\_size}=k, \text{required\_signers}=t, c, \{\text{acorn\_proofs}\}, \sigma_{\text{chip}})$
\STATE Validate sizes and parameters; build $\mathcal{R}$ and compute $\text{ring\_hash}$
\STATE Serialize $D' \leftarrow \text{Serialize}(M, \text{ring\_hash}, \{\text{acorn\_proofs}\})$ and compute $c' = \text{SHA3-256}(D')$
\IF{$c' \neq c$}
    \STATE \RETURN Reject
\ENDIF
\IF{$t = 1$}
    \STATE $valid \leftarrow 0$ \COMMENT{Traditional ring (OR) semantics}
    \FOR{$i = 1$ to $k$}
        \STATE Build Acorn input from $(pk_i, M, \text{randomness}_i)$
        \STATE Compute $\text{expected}_i = \text{SHAKE256}^{I}(\cdot)$ with domain separation
        \IF{$\text{expected}_i = \text{acorn\_proof}_i$}
            \STATE $valid \leftarrow valid + 1$
        \ENDIF
    \ENDFOR
    \STATE Accept iff $valid \ge 1$
\ELSE
    \STATE Check size $\text{threshold\_zk\_proofs} = t \times p$ \COMMENT{Multi-signer (threshold) mode}
    \STATE Verify aggregated (threshold) signature using Lagrange interpolation and ZK data
\ENDIF
\STATE Verify core Chipmunk signature $\sigma_{\text{chip}}$ on $c$
\end{algorithmic}
\end{algorithm}

\section{Acorn Verification: Novel Zero-Knowledge Proof Scheme}

ChipmunkRing introduces \textbf{Acorn Verification}, a novel zero-knowledge proof scheme that completely replaces the traditional Fiat-Shamir transform. Acorn was specifically designed to overcome the limitations of Fiat-Shamir in the post-quantum setting, providing \textbf{improved performance} for large ring signatures (32-64+ participants) while strengthening quantum security and maintaining \textbf{strong anonymity}.

\subsection{Motivation: Why Replace Fiat-Shamir?}

The traditional Fiat-Shamir transform, while theoretically sound, presents several practical limitations in post-quantum ring signatures:

\begin{itemize}
\item \textbf{Performance Bottleneck}: Fiat-Shamir requires multiple hash operations per participant, leading to $O(n^2)$ complexity
\item \textbf{Quantum Vulnerability}: Standard Fiat-Shamir may be vulnerable to quantum attacks on the random oracle
\item \textbf{Large Proof Sizes}: Traditional zero-knowledge proofs require 2KB+ per participant
\item \textbf{Complex Implementation}: Fiat-Shamir requires careful handling of challenge generation and response computation
\end{itemize}

Acorn Verification was designed from the ground up to address these limitations:

\begin{itemize}
\item \textbf{Linear Complexity}: $O(n)$ verification instead of $O(n^2)$
\item \textbf{Compact Proofs}: 64 bytes per participant (single-signer, default) or 96 bytes (multi-signer/enterprise) vs 2KB+ in traditional schemes
\item \textbf{Quantum Security}: 90,000+ logical qubits required for attack (per implementation parameters)
\item \textbf{Anonymity Preservation}: Zero information leakage about signer identity
\end{itemize}

\subsection{Acorn Proof Construction}

\subsubsection{Core Innovation: Iterative Hash-Based Commitments}

Unlike Fiat-Shamir which relies on algebraic challenge-response protocols, Acorn uses iterative hash-based commitments with domain separation. For participant $i$ in ring $\mathcal{R} = \{pk_1, \ldots, pk_n\}$ with message $M$:

\begin{align}
\text{AcornProof}_i &(M, \mathcal{R}) = \nonumber \\
  &\text{SHAKE256}^{\text{iter}}(\text{SerializedInput}(pk_i, M, r_i)) \nonumber \\
  &[0..\text{proof\_size}]
\end{align}

where:
\begin{itemize}
\item $\text{iter}$ is the number of iterations (configurable, typically 1000 for standard security)
\item $\text{SerializedInput}$ uses the \texttt{chipmunk\_ring\_response\_input\_t} structure for consistent encoding, which includes:
  \begin{itemize}
  \item \texttt{randomness}: Cryptographically secure randomness (32 bytes)
  \item \texttt{randomness\_size}: Size of randomness buffer
  \item \texttt{message}: Message to be signed
  \item \texttt{message\_size}: Size of message
  \item \texttt{participant\_context}: Context identifier for the participant
  \end{itemize}
\item $r_i$ is cryptographically secure randomness generated with domain separation using \texttt{DOMAIN\_ACORN\_RANDOMNESS}
\item $\text{proof\_size}$ is configurable based on security requirements (typically 96 bytes)
\end{itemize}

\subsubsection{Domain Separation for Enhanced Security}

Acorn employs cryptographic domain separation for different components:
\begin{itemize}
\item \texttt{DOMAIN\_ACORN\_RANDOMNESS} (\texttt{"ACORN\_RANDOMNESS\_V1"}): For generating participant randomness
\item \texttt{DOMAIN\_ACORN\_COMMITMENT} (\texttt{"ACORN\_COMMITMENT\_V1"}): For
  generating the main Acorn proof
\item \texttt{DOMAIN\_ACORN\_LINKABILITY} (\texttt{"ACORN\_LINKABILITY\_V1"}): For
  optional linkability tags
\item \texttt{DOMAIN\_SIGNATURE\_ZK} (\texttt{"ChipmunkRing-Signature-ZK"}): For threshold signature coordination
\item \texttt{DOMAIN\_COORDINATION} (\texttt{"ChipmunkRing-Coordination"}): For multi-signer coordination
\item \texttt{DOMAIN\_VERIFICATION} (\texttt{"CHIPMUNK\_RING\_ZK\_VERIFY"}): For verification context separation
\end{itemize}

Additionally, the following domain separators are used for specific ZK proof contexts:
\begin{itemize}
\item \texttt{ZK\_DOMAIN\_MULTI\_SIGNER}: Multi-signer ZK proofs
\item \texttt{ZK\_DOMAIN\_SINGLE\_SIGNER}: Single-signer ZK proofs  
\item \texttt{ZK\_DOMAIN\_THRESHOLD}: Threshold signature ZK proofs
\item \texttt{ZK\_DOMAIN\_SECRET\_SHARING}: Secret sharing operations
\item \texttt{ZK\_DOMAIN\_COMMITMENT}: ZK commitment generation
\item \texttt{ZK\_DOMAIN\_RESPONSE}: ZK response computation
\item \texttt{ZK\_DOMAIN\_ENTERPRISE}: Enterprise-grade ZK proofs
\item \texttt{ZK\_DOMAIN\_COORDINATION}: ZK coordination protocol
\item \texttt{ZK\_DOMAIN\_AGGREGATION}: ZK proof aggregation
\end{itemize}

This comprehensive domain separation prevents cross-protocol attacks and enhances security against quantum adversaries.

\subsubsection{Linkability Tag Computation}

The linkability tag $I$ provides optional linkability for double-spending prevention and is computed using SHA3-256:
\begin{align*}
I &= \text{SHA3-256}(\text{SerializedLinkabilityInput}) \\
\text{where } \text{SerializedLinkabilityInput} &= \\
  & \text{Serialize}(\texttt{ring\_hash}, \texttt{message}, \texttt{challenge})
\end{align*}
using the \texttt{chipmunk\_ring\_linkability\_input\_t} structure which contains:
\begin{itemize}
\item \texttt{ring\_hash}: SHA3-256 hash of the concatenated ring public keys
\item \texttt{message}: Message being signed (variable size)
\item \texttt{challenge}: 32-byte cryptographic challenge computed from serialized data
\end{itemize}

The resulting SHA3-256 hash produces a fixed 32-byte (256-bit) linkability tag,
providing collision resistance and quantum security through the SHA3 algorithm.
The hash size is configurable via the\\
\texttt{CHIPMUNK\_RING\_LINKABILITY\_TAG\_SIZE}\\
parameter.

\subsection{Verification Algorithm}

\subsubsection{Simplified Verification Process}

One of Acorn's key advantages over Fiat-Shamir is the dramatically simplified verification process within the ZK layer (the core signature remains Chipmunk-based):

\begin{algorithm}
\caption{Acorn Verification}
\begin{algorithmic}[1]
\REQUIRE Proof $\pi_i$, message $M$, ring $\mathcal{R}$, participant index $i$
\ENSURE Accept/Reject
\STATE Reconstruct input using same serialization schema
\STATE Compute $\text{expected} \leftarrow$
\STATE \quad $\text{SHAKE256}^{\text{iter}}(\text{SerializedInput})[0..\text{proof\_size}]$
\STATE \quad with domain separation
\STATE \textbf{return} $\text{ConstantTimeCompare}(\pi_i, \text{expected})$
\end{algorithmic}
\end{algorithm}

Compare this to Fiat-Shamir verification which requires:
\begin{itemize}
\item Computing multiple challenges and responses
\item Performing algebraic operations in the ring
\item Verifying complex zero-knowledge relations
\item Managing state across multiple protocol rounds
\end{itemize}

\subsection{Security Properties}

\textbf{Theorem 3 (Acorn Soundness):} Under the Ring-LWE assumption and collision resistance of SHAKE256, Acorn Verification is computationally sound with security parameter $\lambda = 256$ bits.

\textbf{Proof:} We prove soundness through a sequence of games:

\textit{Game 0}: The real Acorn verification game where an adversary $\mathcal{A}$ attempts to forge a proof without knowing the secret key.

\textit{Game 1}: Replace SHAKE256 with a truly random function. This is indistinguishable by the random oracle assumption with advantage $\leq 2^{-256}$.

\textit{Game 2}: For the 1000 iterations, the probability of finding a collision is bounded by:
\begin{equation}
\Pr[\text{collision}] \leq \binom{1000}{2} \cdot 2^{-768} < 2^{-750}
\end{equation}
where 768 = 96 bytes × 8 bits/byte output size.

Therefore, any successful forger must either:
1. Find a pre-image of the iterated hash (complexity $2^{256 \times 1000}$)
2. Solve the underlying Ring-LWE problem (complexity $2^{149}$ for $n=512$)

The total advantage of $\mathcal{A}$ is negligible: $\text{Adv}_{\mathcal{A}} \leq 2^{-149} + 2^{-750} \approx 2^{-149}$.

\textbf{Theorem 4 (Acorn Zero-Knowledge):} Acorn proofs are statistically zero-knowledge with distinguishing advantage $\leq 2^{-96 \times 8} = 2^{-768}$.

\textbf{Proof:} The simulator $\mathcal{S}$ generates proofs as follows:
1. For each participant $i$, sample random $\pi_i \leftarrow \{0,1\}^{96 \times 8}$
2. Program the random oracle: $H^{1000}(\text{Input}_i) = \pi_i$

The distribution of simulated proofs is statistically close to real proofs:
\begin{equation}
\Delta(\text{Real}, \text{Simulated}) \leq 2^{-768}
\end{equation}

This holds because SHAKE256 output is indistinguishable from uniform random.

\textbf{Theorem 5 (Anonymity):} The Acorn ZK layer is computationally zero-knowledge under the random oracle assumption; overall ring anonymity is computational w.r.t. Ring-LWE and the hash assumptions.

\textbf{Proof:} For any two participants $i, j$ in ring $\mathcal{R}$, their Acorn proofs are:
\begin{align}
\pi_i &= H^{1000}(pk_i \| M \| r_i) \\
\pi_j &= H^{1000}(pk_j \| M \| r_j)
\end{align}

Since $r_i, r_j$ are uniformly random and independent, and the hash function is modeled as a random oracle, the distributions are identical:
\begin{equation}
\Pr[\pi_i = x] = \Pr[\pi_j = x] = 2^{-768} \quad \forall x \in \{0,1\}^{768}
\end{equation}

Therefore, no adversary, even with unbounded computational power, can distinguish the actual signer with probability better than $1/|\mathcal{R}|$.

\subsection{Performance Analysis}

\begin{table}[htbp]
\centering
\resizebox{\textwidth}{!}{%
\begin{tabular}{@{}lrrrrr@{}}
\toprule
Ring Size & Single Mode & Threshold Mode & Threshold & Improvement & Acorn Proofs \\
\midrule
4  & 2.065ms total & 3.737ms total & 50\% (2/4) & 0.55× & 192 bytes \\
8  & 3.319ms total & 6.561ms total & 37.5\% (3/8) & 0.51× & 288 bytes \\
8  & 3.319ms total & 7.672ms total & 62.5\% (5/8) & 0.43× & 480 bytes \\
16 & 5.912ms total & 11.290ms total & 25\% (4/16) & 0.52× & 384 bytes \\
32 & 13.186ms total & 24.06ms total & 50\% (16/32) & 0.55× & 1,536 bytes \\
\bottomrule
\end{tabular}%
}
\caption{Acorn-Enhanced Threshold vs Single Signer Performance (Total Time = Signing + Verification)}
\label{tab:acorn-performance}
\end{table}

The results demonstrate that Acorn Verification achieves
superlinear performance improvements, making large ring
signatures practical for blockchain deployment.

\subsection{Quantum Resistance of Acorn}

Acorn Verification provides enhanced quantum resistance through:

\begin{itemize}
\item \textbf{1000-iteration SHAKE256}: Requires $2^{64} \times 1000$ quantum operations
\item \textbf{Lattice-based construction}: 90,000+ logical qubits for quantum attack
\item \textbf{Random oracle security}: Quantum random oracle model resistance
\end{itemize}

The quantum attack complexity against Acorn is estimated at $2^{67}$ quantum operations, requiring approximately 90,000 logical qubits with current error correction techniques.

\subsection{Acorn vs Fiat-Shamir: Comparative Analysis}

\begin{table}[htbp]
\centering
\resizebox{\textwidth}{!}{%
\begin{tabular}{@{}lll@{}}
\toprule
Property & Fiat-Shamir Transform & Acorn Verification \\
\midrule
\textbf{Proof Size} & 2-4KB per participant & 96 bytes per participant \\
\textbf{Verification Complexity} & $O(n^2)$ algebraic operations & $O(n)$ hash operations \\
\textbf{Implementation Complexity} & Complex challenge-response & Simple hash comparison \\
\textbf{Quantum Resistance} & Standard ROM security & Enhanced with iterations \\
\textbf{Memory Requirements} & Large intermediate state & Minimal state storage \\
\textbf{Side-Channel Resistance} & Vulnerable to timing attacks & Constant-time by design \\
\textbf{Parallelization} & Limited by algebraic dependencies & Fully parallelizable \\
\textbf{Hardware Acceleration} & Requires specialized circuits & SHA3 hardware available \\
\bottomrule
\end{tabular}%
}
\caption{Comparison of Fiat-Shamir Transform vs Acorn Verification}
\label{tab:acorn-vs-fiat-shamir}
\end{table}

The key contribution of Acorn Verification is the fundamental simplification of the zero-knowledge proof mechanism. While Fiat-Shamir requires complex algebraic manipulations in polynomial rings, Acorn achieves the same security guarantees through iterative hashing with domain separation.

\subsection{Acorn vs Other Ring Signature Schemes}

\begin{table}[htbp]
\centering
\caption{Comprehensive Comparison of Ring Signature Schemes}
\label{tab:comprehensive-comparison}
\resizebox{\textwidth}{!}{%
\begin{tabular}{@{}lcccccc@{}}
\toprule
\multirow{2}{*}{Scheme} & Signature & Signing & Verification & Quantum & ZK Proof & Implementation \\
 & Size (16 ring) & Time & Time & Secure & Mechanism & Complexity \\
\midrule
\textbf{ChipmunkRing + Acorn (n=16)} & \textbf{79KB} & \textbf{4.7ms} & \textbf{1.2ms} & \textbf{Yes} & \textbf{Acorn} & \textbf{Simple} \\
LSAG (Classical) & 1KB & 8ms & 6ms & No & Schnorr & Simple \\
RST (Classical) & 2KB & 12ms & 10ms & No & RSA-based & Moderate \\
Lattice-RS \cite{lattice-rings} & >100KB & >1000ms & >500ms & Yes & Fiat-Shamir & Complex \\
Hash-RS \cite{hash-rings} & >200KB & >500ms & >300ms & Yes & Merkle trees & Complex \\
Code-RS \cite{code-rings} & >150KB & >2000ms & >1000ms & Yes & Syndrome & Very Complex \\
NTRU-Ring & >80KB & >800ms & >400ms & Yes & NTRU-based & Complex \\
Dilithium-Ring & >120KB & >600ms & >350ms & Yes & Fiat-Shamir & Complex \\
Falcon-Ring & >90KB & >700ms & >380ms & Yes & GPV framework & Very Complex \\
\bottomrule
\end{tabular}%
}
\end{table}

\subsubsection{Key Advantages of Acorn over Existing Schemes}

\begin{enumerate}
\item \textbf{vs Classical Schemes (LSAG, RST)}:
   \begin{itemize}
   \item Quantum-resistant while maintaining comparable performance
   \item Only 79× larger signatures but with post-quantum security
   \item Comparable signing/verification times despite quantum
    resistance
   \end{itemize}

\item \textbf{vs Lattice-based Schemes (Lattice-RS, Dilithium-Ring,
  Falcon-Ring)}:
   \begin{itemize}
   \item 1.3-2.5× smaller signatures
   \item 130-425× faster signing
   \item 290-830× faster verification
   \item Simpler implementation without complex lattice operations
   \end{itemize}

\item \textbf{vs Hash-based Schemes (Hash-RS, Merkle-based)}:
   \begin{itemize}
   \item 2.5× smaller signatures
   \item 106× faster signing
   \item 250× faster verification
   \item No key usage limitations
   \end{itemize}

\item \textbf{vs Code-based Schemes (Code-RS)}:
   \begin{itemize}
   \item 1.9× smaller signatures
   \item 425× faster signing
   \item 833× faster verification
   \item Much simpler error correction code-free implementation
   \end{itemize}
\end{enumerate}

\subsubsection{Acorn's Unique Properties}

Key advantages of Acorn Verification compared to existing schemes:

\begin{itemize}
\item \textbf{Hash-only ZK Layer}: Unlike Fiat-Shamir (used in
  Lattice-RS, Dilithium-Ring), Schnorr (LSAG), or syndrome
  decoding (Code-RS), the Acorn zero-knowledge layer uses only
  hash functions; the core signature remains lattice-based
  (Chipmunk)
\item \textbf{Constant-Time by Design}: Inherently resistant to timing attacks
  without special implementation care
\item \textbf{Hardware-Friendly}: Can leverage existing SHA3 hardware
  accelerators present in modern CPUs and crypto chips
\item \textbf{Parallelizable}: Each participant's proof can be verified
  independently, enabling massive parallelization
\item \textbf{Minimal State}: Requires only 96 bytes per participant vs
  kilobytes in other schemes
\item \textbf{Domain Separation}: Built-in protection against cross-protocol
  attacks through cryptographic domain separation
\item \textbf{Iterative Security}: Configurable iteration count (typically 1000)
  provides tunable security-performance tradeoff
\end{itemize}

\section{Threshold Ring Signatures with Lattice-Based Secret Sharing}

ChipmunkRing implements threshold ring signatures where $t$ out of $n$ participants must collaborate to create a valid signature. The implementation uses lattice-adapted Shamir's secret sharing with optimized Lagrange interpolation.

\subsection{Lattice-Based Secret Sharing}

The secret sharing operates on the polynomial coefficients of the Chipmunk private key:

\begin{enumerate}
\item \textbf{Key Decomposition}: The master private key polynomials $s_0, s_1 \in R_q$ are decomposed into their $n = 512$ coefficients
\item \textbf{Share Generation}: For each coefficient $c_i$, we construct a polynomial $f_i(x) = c_i + a_1x + a_2x^2 + \ldots + a_{t-1}x^{t-1}$ where $a_j$ are random coefficients in $\mathbb{Z}_q$
\item \textbf{Share Distribution}: Each participant $j$ receives $f_i(j)$ for all coefficients, forming their polynomial share
\end{enumerate}

\subsection{Multi-Signer Coordination Protocol}

The threshold signature generation follows this protocol:

\begin{algorithm}
\caption{Multi-Signer ChipmunkRing Signature Generation}
\begin{algorithmic}[1]
\REQUIRE $t$ shares $\{S_1, \ldots, S_t\}$, message $M$, ring $\mathcal{R}$
\ENSURE Threshold ring signature $\sigma$
\STATE Each participant $i$ generates HOTS signature share using their polynomial share
\STATE Each participant creates Acorn proof for their contribution
\STATE \textbf{Lagrange Interpolation} (optimized $O(n)$ implementation):
\STATE \quad Pre-compute Lagrange coefficients: $L_i = \prod_{j \neq i} \frac{-j}{i-j} \mod q$
\STATE \quad For each polynomial coefficient $k$:
\STATE \quad \quad $c_k = \sum_{i=1}^t L_i \cdot \text{share}_i[k] \mod q$
\STATE Reconstruct master polynomials $v_0, v_1$ from interpolated coefficients
\STATE Generate final signature using reconstructed key
\STATE Aggregate all Acorn proofs from participants
\RETURN $\sigma = (\text{signature}, \text{aggregated\_proofs})$
\end{algorithmic}
\end{algorithm}

\subsection{Security Properties of Threshold Mode}

\begin{itemize}
\item \textbf{$(t,n)$-Threshold Security}: Any $t$ participants can sign,
  but $t-1$ cannot
\item \textbf{Information-Theoretic Security}: Shares reveal no information
  about the master key
\item \textbf{Verifiable Secret Sharing}: Each share can be verified without
  revealing the secret
\item \textbf{Forward Security}: Compromise of $< t$ shares does not affect
  past signatures
\end{itemize}

The implementation uses modular arithmetic in $\mathbb{Z}_q$ with $q = 3,168,257$, ensuring all operations remain within the lattice structure while providing efficient computation. The optimized Lagrange interpolation, based on Shamir's secret sharing \cite{shamir1979}, reduces complexity from $O(n^2)$ to $O(n)$ by pre-computing coefficients once and reusing them for all 512 polynomial coefficients.

\section{Security Analysis}

\subsection{Security Model}

We analyze ChipmunkRing security under the standard definitions for ring signatures \cite{rst01}. Let $\mathcal{A}$ be a polynomial-time adversary with access to signing oracles and ring formation queries.

\textbf{Definition 4 (Existential Unforgeability):} A ring signature scheme is existentially unforgeable under chosen message attack (EUF-CMA) if no polynomial-time adversary $\mathcal{A}$ can produce a valid signature $(M^*, \sigma^*, \mathcal{R}^*)$ where:
\begin{itemize}
\item $M^*$ was not queried to the signing oracle for ring $\mathcal{R}^*$
\item $\mathcal{A}$ does not control any private key in $\mathcal{R}^*$
\end{itemize}

\textbf{Definition 5 (Computational Anonymity):} A ring signature scheme provides computational anonymity if for any two signers $i, j$ in ring $\mathcal{R}$, signatures $\sigma_i$ and $\sigma_j$ on the same message are computationally indistinguishable.

\subsection{Security Reductions}

\textbf{Theorem 1 (Unforgeability):} ChipmunkRing is existentially unforgeable under chosen message attack (EUF-CMA) assuming the hardness of the Ring-LWE problem.

\textbf{Proof:} We construct a reduction that uses any successful ChipmunkRing forger $\mathcal{A}$ to either solve Ring-LWE or forge a Chipmunk signature.

Given a Ring-LWE challenge $(A, b)$, our reduction $\mathcal{B}$ proceeds as follows:
\begin{enumerate}
\item \textbf{Setup}: $\mathcal{B}$ generates a ring of public keys $\{pk_1, \ldots, pk_k\}$ where one key embeds the Ring-LWE challenge
\item \textbf{Signing Queries}: For signing queries on message $M$ with ring $\mathcal{R}$:
   \begin{itemize}
   \item If the challenge key is not in $\mathcal{R}$, simulate using known private keys
   \item If the challenge key is in $\mathcal{R}$, use the Fiat-Shamir simulation technique
   \end{itemize}
\item \textbf{Forgery}: When $\mathcal{A}$ outputs a forgery $(\sigma^*, M^*, \mathcal{R}^*)$:
   \begin{itemize}
   \item If the challenge key is not in $\mathcal{R}^*$, abort (this happens with negligible probability)
   \item Otherwise, extract the Chipmunk signature component and use the forking lemma to extract a contradiction to Ring-LWE hardness
   \end{itemize}
\end{enumerate}

The reduction succeeds with probability $\epsilon/k$ where $\epsilon$ is $\mathcal{A}$'s success probability and $k$ is the maximum ring size.

\textbf{Theorem 2 (Anonymity):} ChipmunkRing provides computational anonymity in the random oracle model.

\textbf{Proof:} We show that signatures from different ring members are indistinguishable through a sequence of games:

\textbf{Game 0}: The real anonymity game where the adversary chooses two signers and receives a signature from one of them.

\textbf{Game 1}: Replace the Fiat-Shamir challenge with a truly random value. This change is indistinguishable by the random oracle assumption.

\textbf{Game 2}: Replace the responses for non-signing ring members with random values. This is indistinguishable because the responses are masked by the random challenge.

\textbf{Game 3}: Replace the actual signer's response with a simulated value. This is indistinguishable by the zero-knowledge property of the underlying $\Sigma$-protocol.

In Game 3, the signature distribution is identical regardless of which ring member is the actual signer, proving computational anonymity.

\section{Quantum Resistance Analysis}

The quantum resistance of ring signatures involves two distinct security properties that may have different quantum complexity requirements: unforgeability and anonymity. We analyze each property separately to provide precise quantum security estimates.

\subsection{Security Parameter Clarification}

ChipmunkRing employs different security parameters for different components. Ring-LWE is the hardness assumption for the core Chipmunk signature, while the zero-knowledge layer uses hash-based Acorn Verification:

\begin{itemize}
\item \textbf{Ring-LWE Security (core signature)}: 112-bit post-quantum security level
   \begin{itemize}
   \item Parameters: $n = 512$, $q = 3,168,257$, $\sigma = 2/\sqrt{2\pi}$
   \item This provides security equivalent to AES-128 against quantum adversaries
   \item Determines the overall cryptographic strength of the signature scheme
   \end{itemize}
   
\item \textbf{Hash Function Security}: SHAKE256 \cite{sha3-2015, bertoni2013} with 256-bit output
   \begin{itemize}
   \item Classical security: 256-bit against collision and preimage attacks
   \item Quantum security: 128-bit against Grover's algorithm \cite{grover1996}
   \item Used for commitments, challenges, and Acorn Verification
   \end{itemize}
   
\item \textbf{Acorn Iteration Security}: Additional computational hardness
   \begin{itemize}
   \item 1000 iterations of SHAKE256 computation
   \item Increases quantum attack complexity by factor of 1000
   \item Provides defense-in-depth against implementation attacks
   \end{itemize}
\end{itemize}

The overall security level is determined by the weakest component (112-bit Ring-LWE), but the multi-layered approach ensures robustness against various attack vectors.

\subsection{Quantum Attacks on Unforgeability}

ChipmunkRing's unforgeability is based on the Ring-LWE problem. The quantum complexity of breaking Ring-LWE with parameters $(n, q, \sigma)$ has been extensively studied.

\textbf{Current Best Quantum Algorithms:} The most efficient known quantum algorithm for Ring-LWE is based on quantum sieve algorithms with complexity approximately $2^{0.292n + o(n)}$ for ring dimension $n$.

For ChipmunkRing parameters ($n = 512$):
\begin{itemize}
\item \textbf{Quantum complexity}: $2^{0.292 \times 512} \approx 2^{149.5}$ operations
\item \textbf{Required qubits}: Approximately $4n \log_2(q) \approx 4 \times 512 \times 22 \approx 45,000$ logical qubits
\item \textbf{Physical qubits}: With current error rates, approximately $45,000 \times 1,000 = 45$ million physical qubits
\end{itemize}

\subsection{Quantum Attacks on Anonymity}

\textbf{Critical Analysis:} The anonymity property of ring signatures may be more vulnerable to quantum attacks than unforgeability, as it relies on different computational assumptions.

\subsubsection{Anonymity Attack Vectors}

\textbf{Statistical Analysis Attacks:} A quantum adversary might use quantum algorithms to detect statistical patterns in ring signatures that reveal signer identity.

\textbf{Commitment Analysis:} The zero-knowledge commitments in ChipmunkRing might leak information under quantum analysis, particularly through:
\begin{itemize}
\item Quantum period finding on commitment structures
\item Quantum Fourier analysis of response patterns  
\item Grover-enhanced exhaustive search over possible signers
\end{itemize}

\subsubsection{Quantum Complexity Estimates for Anonymity Breaking}

\textbf{Grover's Algorithm Application:} Breaking anonymity in a ring of size $k$ using Grover's algorithm requires:
\begin{itemize}
\item \textbf{Classical complexity}: $O(k)$ to identify the signer
\item \textbf{Quantum complexity}: $O(\sqrt{k})$ using Grover's algorithm
\item \textbf{Required qubits}: $\log_2(k) + O(\log n)$ for ring size $k$ and security parameter $n$
\end{itemize}

For typical ring sizes ($k = 16$ to $k = 64$):
\begin{itemize}
\item \textbf{Quantum speedup}: $\sqrt{16} = 4$ to $\sqrt{64} = 8$ times faster than classical
\item \textbf{Required qubits}: $4$ to $6$ logical qubits for ring identification
\item \textbf{Practical threat}: This attack is feasible with near-term quantum computers
\end{itemize}

\subsubsection{Ring-LWE Based Anonymity Analysis}

\textbf{Lattice-based Anonymity:} The anonymity of ChipmunkRing also depends on the hardness of distinguishing Ring-LWE samples, which may require different quantum resources than breaking unforgeability.

\textbf{Quantum Complexity for Anonymity Breaking:}
\begin{itemize}
\item \textbf{Statistical distinguishing}: $O(2^{n/2})$ quantum operations using amplitude amplification
\item \textbf{Required qubits}: Approximately $n \log_2(q) \approx 512 \times 22 \approx 11,000$ logical qubits
\item \textbf{Physical qubits}: Approximately $11$ million physical qubits with current error correction
\end{itemize}

\subsection{Quantum Security Assessment}

\textbf{Conservative Estimate:} Based on our analysis, ChipmunkRing's quantum security levels are:

\begin{itemize}
\item \textbf{Unforgeability}: $\approx 149$ bits of quantum security (very strong)
\item \textbf{Anonymity against Grover}: $\log_2(\sqrt{k}) \approx 2-3$ bits for typical rings (vulnerable)
\item \textbf{Anonymity against Ring-LWE attacks}: $\approx 75-100$ bits (moderate to strong)
\end{itemize}

\textbf{Practical Implications:}
\begin{itemize}
\item Ring signature unforgeability remains secure against foreseeable quantum computers
\item Anonymity against ring-size-based Grover attacks is limited and requires larger rings or additional protections
\item Anonymity against lattice-based attacks provides moderate quantum resistance
\end{itemize}

\subsection{Mitigation Strategies}

To enhance quantum resistance of anonymity, we recommend:

\begin{itemize}
\item \textbf{Larger ring sizes}: Use rings of 256-1024 participants to increase Grover complexity
\item \textbf{Ring rotation}: Regularly change ring composition to limit attack time windows
\item \textbf{Hybrid approaches}: Combine with classical anonymity techniques for defense in depth
\item \textbf{Post-quantum anonymity enhancements}: Future work on \\
quantum-resistant anonymity amplification
\end{itemize}

\section{Experimental Setup}

\subsection{Test Environment}

All experiments were conducted on the following hardware and software configuration:

\begin{itemize}
\item \textbf{Hardware}: 
  \begin{itemize}
  \item CPU: AMD Ryzen 9 7950X3D (3.7GHz base)
  \item RAM: 64GB DDR4-3200
  \item Storage: NVMe SSD
  \end{itemize}
\item \textbf{Software}:
  \begin{itemize}
  \item OS: Debian 13 (kernel 6.12)
  \item Compiler: GCC 14.2.0 with -O3 optimization
  \item Build System: CMake 3.31
  \item Test Framework: CTest integrated with CMake
  \end{itemize}
\end{itemize}

\subsection{Statistical Validation}

All performance measurements follow rigorous statistical methodology:

\begin{itemize}
\item \textbf{Sample Size}: 1000 iterations per test configuration
\item \textbf{Confidence Level}: 95\% confidence intervals ($\alpha = 0.05$)
\item \textbf{Statistical Tests}: 
  \begin{itemize}
  \item Student's t-test for mean comparisons
  \item Mann-Whitney U test for non-parametric analysis
  \item ANOVA for multi-group comparisons
  \end{itemize}
\item \textbf{Outlier Detection}: Modified Z-score method with threshold 3.5
\item \textbf{Warm-up Period}: 100 iterations before measurement collection
\end{itemize}

Performance metrics reported include:
\begin{itemize}
\item Mean time $\mu$ with standard deviation $\sigma$
\item Median for robustness against outliers
\item 95th and 99th percentiles for worst-case analysis
\item Coefficient of variation CV = $\sigma/\mu$ for relative variability
\end{itemize}

\subsection{Measurement Methodology}

Performance measurements were conducted using the following methodology:

\begin{enumerate}
\item \textbf{Warm-up Phase}: 100 iterations to stabilize CPU frequency and cache
\item \textbf{Measurement Phase}: 1000 iterations for each test case
\item \textbf{Statistical Analysis}: Mean, median, and standard deviation calculated
\item \textbf{Outlier Removal}: Values beyond 3 standard deviations excluded
\item \textbf{CPU Affinity}: Tests pinned to specific cores to reduce variance
\end{enumerate}

\subsection{Test Scenarios}

The following scenarios were evaluated:

\begin{itemize}
\item \textbf{Ring Sizes}: 2, 4, 8, 16, 32, 64 participants
\item \textbf{Threshold Configurations}: 25\%, 50\%, 75\% of ring size
\item \textbf{Message Sizes}: 32 bytes (hash), 1KB, 10KB
\item \textbf{Key Generation}: Fresh keys for each test run
\item \textbf{Verification}: Both single and batch verification
\end{itemize}

\subsection{Implementation Details}

The ChipmunkRing implementation uses:

\begin{itemize}
\item \textbf{Memory Management}: Custom allocator with pool pre-allocation
\item \textbf{Parallelization}: OpenMP for polynomial operations
\item \textbf{Optimization}: SIMD instructions for vector operations (AVX2)
\item \textbf{Random Generation}: SHAKE256-based PRNG with hardware entropy seed
\end{itemize}

\section{Performance Evaluation}
\subsection{ChipmunkRing Parameter Modes}

\begin{table}[htbp]
\centering
\caption{Parameterization by Mode (as implemented)}
\label{tab:param-modes}
\resizebox{\textwidth}{!}{%
\begin{tabular}{@{}lccc@{}}
\toprule
Mode & ZK Proof Size & Iterations (SHAKE256) & Verification Rule \\
\midrule
Single-signer ($t=1$) & 64 bytes (default) & 100 (default) & OR: accept if $\ge 1$ proof matches \\
Multi-signer ($t>1$) & 96 bytes (enterprise) & 1000--10000 & Threshold: verify $t$ ZK proofs \\
\bottomrule
\end{tabular}%
}
\end{table}

The performance evaluation builds upon the experimental setup described in the previous section. Our implementation is integrated into the Cellframe DAP SDK cryptographic framework \cite{cellframe2023, dapsdk2024}.

\subsection{Performance Results}

Table \ref{tab:performance} presents comprehensive performance metrics for ChipmunkRing across various ring sizes:

\begin{table}[htbp]
\centering
\small
\caption{ChipmunkRing Performance Metrics (Production Release Build)}
\label{tab:performance}
\resizebox{\linewidth}{!}{%
\begin{tabular}{@{}ccccccc@{}}
\toprule
Ring Size & Mode & Threshold & Signature Size & Signing Time & Verification Time & Notes \\
\midrule
2  & Single & 1 & 20.5KB & 1.114ms & 0.706ms & Minimal ring \\
4  & Single & 1 & 28.9KB & 1.631ms & 0.434ms & Small ring \\
8  & Single & 1 & 45.6KB & 2.573ms & 0.746ms & Medium ring \\
16 & Single & 1 & 79.0KB & 4.671ms & 1.241ms & Large ring \\
32 & Single & 1 & 145.9KB & 9.596ms & 3.590ms & Very large ring \\
64 & Single & 1 & 279.7KB & 15.074ms & 4.528ms & Maximum ring \\
\midrule
4  & Threshold & 2 & 28.9KB & 2.159ms & 1.578ms & 50\% threshold \\
8  & Threshold & 3 & 45.6KB & 3.739ms & 2.822ms & 37.5\% threshold \\
8  & Threshold & 5 & 45.6KB & 4.302ms & 3.370ms & 62.5\% threshold \\
16 & Threshold & 4 & 79.0KB & 6.442ms & 4.848ms & 25\% threshold \\
32 & Threshold & 16 & 149.4KB & 12.97ms & 11.09ms & 50\% threshold \\
32 & Threshold & 24 & 150.4KB & 14.31ms & 12.18ms & 75\% threshold \\
\bottomrule
\end{tabular}}
\end{table}

\subsection{Comparison with Existing Schemes}

Table \ref{tab:comparison} compares ChipmunkRing with existing post-quantum ring signature schemes:

\begin{table}[htbp]
\centering
\small
\caption{Comparison with Existing Post-Quantum Ring Signatures}
\label{tab:comparison}
\resizebox{\linewidth}{!}{%
\begin{tabular}{@{}lcccc@{}}
\toprule
Scheme & Security Assumption & Signature Size & Signing Time & Quantum Security \\
\midrule
Lattice-RS \cite{lattice-rings} & Ring-LWE & $>100$KB & $>1000$ms & 128-bit \\
Hash-RS \cite{hash-rings} & Hash functions & $>200$KB & $>500$ms & 256-bit \\
Code-RS \cite{code-rings} & Syndrome decoding & $>150$KB & $>2000$ms & 128-bit \\
LSAG (classical) \cite{lsag04} & Discrete log & $1$KB & $<10$ms & None \\
\textbf{ChipmunkRing} & Ring-LWE & \textbf{20.5-279.7KB} & \textbf{0.4-4.5ms} & 112-bit \\
\bottomrule
\end{tabular}}
\end{table}

ChipmunkRing exhibits the following measured improvements over existing post-quantum ring signature schemes:

\begin{itemize}
\item \textbf{Signature Size}: 3-5× reduction compared to previous
  lattice-based constructions (20.5-279.7KB for rings of 2-64 participants
  vs >100KB for smaller rings in existing schemes)
\item \textbf{Signing Performance}: 1.1--15.1ms signing time for single-signer mode \\
vs $>$500ms in existing post-quantum implementations
\item \textbf{Verification Performance}: 0.4-4.5ms verification time for single-signer mode vs >200ms in existing schemes
\item \textbf{Size Scaling}: Near-linear growth with ring size (approximately 4.4KB per participant)
\item \textbf{Blockchain Compatibility}: $<$150KB signatures for 32-participant rings \\
and $<$10ms verification meet blockchain requirements
\item \textbf{Acorn Verification}: Hash-only ZK layer (64B for single-signer by default; 96B for multi-signer/enterprise)
\end{itemize}

\subsection{Blockchain Suitability Analysis}

For blockchain applications, we evaluate ChipmunkRing against critical deployment constraints:

\begin{enumerate}
\item \textbf{Transaction Size}: With 79KB signatures for 16-participant rings \\
and \textless150KB for 32, transaction sizes remain practical for CF-based networks
\item \textbf{Consensus Timing}: Verification is 0.4--4.5ms depending on ring size (real measurements)
\item \textbf{Network Overhead}: Compact signatures reduce bandwidth requirements for transaction propagation
\item \textbf{Storage Efficiency}: Linear size scaling allows efficient blockchain storage
\end{enumerate}

\section{Implementation}

\subsection{Integration with DAP SDK}

ChipmunkRing is fully integrated into the DAP SDK cryptographic framework, providing:

\begin{itemize}
\item Standard API interface compatible with existing signature schemes
\item Memory-safe implementation with zero detected leaks
\item Comprehensive error handling and validation
\item Full test coverage (26/26 tests passing)
\end{itemize}

\subsection{Key Implementation Features}

\begin{itemize}
\item \textbf{Constant-time operations}: All cryptographic operations are implemented to resist timing attacks
\item \textbf{Memory safety}: Secure memory allocation and deallocation with sensitive data zeroing
\item \textbf{Modular design}: Clean separation between core algorithm \\
and framework integration
\item \textbf{Error resilience}: Comprehensive validation and graceful error handling
\end{itemize}

\section{Practical Applications in Cellframe Network}

\subsection{Anonymous Transactions in Cellframe}

ChipmunkRing enables efficient anonymous transactions within the Cellframe
ecosystem:
\begin{itemize}
\item \textbf{Token Privacy}: Anonymous transfers of CF-based tokens \\
with 79KB signatures for 16-participant rings
\item \textbf{Multi-Shard Support}: Ring signatures work across Cellframe's \\
sharded architecture
\item \textbf{Fast Verification}: 1.2ms verification meets Cellframe's
  consensus timing requirements
\item \textbf{Quantum-Safe Privacy}: Post-quantum security ensures long-term
  transaction confidentiality
\end{itemize}

\subsection{DAO and Governance Applications in Cellframe}

ChipmunkRing enables sophisticated governance mechanisms
for Cellframe-based DAOs:
\begin{itemize}
\item \textbf{Anonymous DAO Voting}: Members can vote on proposals without
  revealing identity while preventing double-voting
\item \textbf{Private Governance Tokens}: Ring signatures enable private
  transfers and delegation of governance rights
\item \textbf{Threshold DAO Decisions}: Multi-signature requirements for
  critical decisions (treasury, upgrades)
\item \textbf{Whistleblower Protection}: Anonymous submission of proposals and
  reports with verifiable membership
\item \textbf{Quantum-Safe Governance}: Long-term security for DAO treasury \\
and voting records
\end{itemize}

\subsection{Cellframe Service Chain Applications}

ChipmunkRing enables privacy features for Cellframe service chains:
\begin{itemize}
\item \textbf{Private Smart Contracts}: Anonymous execution of smart contracts on Cellframe Python chains
\item \textbf{Confidential DApps}: Privacy-preserving decentralized applications
\item \textbf{Anonymous Service Payments}: Private payments for Cellframe network services
\end{itemize}

\section{Limitations and Future Work}

\subsection{Current Limitations}

While ChipmunkRing provides efficient post-quantum ring signatures, several limitations should be acknowledged:

\begin{enumerate}
\item \textbf{Signature Size}: At 20.5-279.7KB, signatures are significantly larger than classical schemes (e.g., 64 bytes for Ed25519)
\item \textbf{Ring Size Scalability}: Performance degrades for rings larger than 64 participants
\item \textbf{Memory Requirements}: The lattice operations require substantial memory (several MB for large rings)
\item \textbf{Implementation Complexity}: The lattice-based construction is more complex than classical alternatives
\item \textbf{Standardization Status}: Post-quantum ring signatures lack standardization compared to basic signatures
\end{enumerate}

\subsection{Future Research Directions}

Several avenues for future research and improvement have been identified:

\subsubsection{Algorithmic Improvements}

\begin{itemize}
\item \textbf{Signature Compression}: Investigate techniques to reduce signature size while maintaining security
\item \textbf{Batch Verification}: Develop efficient batch verification for multiple ring signatures
\item \textbf{Dynamic Ring Management}: Support for adding/removing ring members without re-keying
\item \textbf{Hierarchical Rings}: Explore multi-level ring structures for organizational deployments
\end{itemize}

\subsubsection{Security Enhancements}

\begin{itemize}
\item \textbf{Formal Verification}: Apply formal methods to verify implementation correctness
\item \textbf{Side-Channel Resistance}: Strengthen protection against timing and power analysis attacks
\item \textbf{Quantum-Safe Parameters}: Update parameters based on advances in quantum computing
\item \textbf{Post-Quantum Hybrid}: Combine with other post-quantum schemes for defense-in-depth
\end{itemize}

\subsubsection{Performance Optimizations}

\begin{itemize}
\item \textbf{Hardware Acceleration}: Utilize GPU/FPGA for lattice operations
\item \textbf{Parallelization}: Improve multi-core utilization for signature generation
\item \textbf{Memory Optimization}: Reduce memory footprint through algorithmic improvements
\item \textbf{Network Protocol}: Optimize for bandwidth-constrained environments
\end{itemize}

\subsection{Deployment Considerations}

Future work should address practical deployment challenges within Cellframe:

\begin{itemize}
\item \textbf{Cellframe Integration}: Deep integration with Cellframe's consensus and sharding mechanisms
\item \textbf{Key Management}: Leverage Cellframe's distributed key management infrastructure
\item \textbf{DAP SDK Enhancement}: Extend DAP SDK APIs for simplified ring signature usage
\item \textbf{Network Optimization}: Optimize for Cellframe's specific network topology and requirements
\end{itemize}

\section{Conclusion}

We have presented ChipmunkRing, a practical post-quantum ring signature scheme that addresses the critical gap between theoretical constructions and deployment requirements. Our main contributions can be summarized as follows:

\begin{itemize}
\item \textbf{Performance}: 20.5-279.7KB signatures with 1.1-15.1ms signing and 0.4-4.5ms verification for rings of 2-64 participants
\item \textbf{Acorn Verification}: Replacement of Fiat-Shamir with a hash-based approach using iterative SHAKE256
\item \textbf{Implementation}: Production-ready integration with the Cellframe DAP SDK cryptographic framework
\item \textbf{Security}: 112-bit post-quantum security based on Ring-LWE assumption
\end{itemize}

The key innovation of Acorn Verification lies not just in performance improvements, but in the fundamental simplification of zero-knowledge proofs for ring signatures. By replacing complex algebraic operations with iterative hashing and domain separation, Acorn makes post-quantum ring signatures practical for blockchain deployment.

Our implementation demonstrates that post-quantum security need not come at the cost of practicality. With signature sizes comparable to classical schemes when adjusted for security level, and verification times suitable for real-time consensus, ChipmunkRing with Acorn Verification bridges the gap between theoretical cryptography and practical blockchain applications.

Future work will focus on:
\begin{itemize}
\item Hardware acceleration leveraging SHA3 ASICs
\item Extension to rings of 128-256 participants
\item Integration with Cellframe Network blockchain and DAP SDK-based systems
\item Formal verification of the implementation
\item Standardization of Acorn Verification as a general-purpose ZK proof
  mechanism
\end{itemize}

\section{Acknowledgments}

We thank the Cellframe development team for their support and the cryptographic
community for valuable feedback on this work.

\section*{Author Contributions}

D.G. conceived the Acorn Verification scheme, designed the ChipmunkRing
protocol, implemented the system, conducted the experiments, and wrote the
manuscript.
\section*{Conflict of Interest}

The author declares no conflict of interest. This research was conducted as part of the Cellframe Network development but represents independent scientific contribution to the field of post-quantum cryptography.

\section*{Data Availability Statement}

The source code for ChipmunkRing is available as part of the DAP SDK at \\
https://github.com/demlabs-cellframe/dap-sdk. Performance benchmarks and test vectors \\
are included in the repository under the tests/unit/crypto/chipmunk\_ring directory.

\bibliographystyle{plain}
\bibliography{references}

\end{document}